\documentclass{elsart}

\usepackage{psfig}

\usepackage{natbib}

\begin{document}

\runauthor{Cicero, Caesar and Vergil}
\def\ion#1#2{#1$\,${\small\rm{#2}}\relax}
%\def\ion#1#2{#1{\small\rm{#2}}\relax}

% -------------------------------------------------------------------------

\begin{frontmatter}

\title{{\it ASCA} Observations of NLS1s}

\author[COL]{Karen M. Leighly}
\address[COL]{Columbia Astrophysics Laboratory, Columbia University, 550
West 120th Street, New York, NY 10025; leighly@astro.columbia.edu}

\begin{abstract}
The study of NLS1s using {\it ASCA} has many advantages.  A representative
sample can be studied; to date, observations of more than 30 NLS1s
have been made.  {\it ASCA} observations are conducted contiguously, so
their X-ray variability properties can be studied systematically.
{\it ASCA} detectors have a broad band pass and moderate energy resolution,
properties which allow their complex X-ray spectrum to be deconvolved.

{\it ASCA} observations have revealed that a soft excess extending up to 1
keV is frequently found in the X-ray spectra from NLS1s.  The hard
X-ray photon index has been shown to be systematically steeper than in
Seyfert 1 galaxies with broad optical lines.  NLS1s also exhibit
larger amplitude variability than broad-line Seyferts of similar hard
X-ray luminosity.  On occasion, features in the soft X-ray band have
been detected that appear to be specific to NLS1s; ionized iron lines
have also been observed.

These results have been fundamental for building our current picture
of the X-ray emission from NLS1s.  Many of the properties can be
explained if the specific accretion rate is larger in NLS1s than in
Seyfert 1 galaxies with broad optical lines.  NLS1s may exemplify an
extreme state of an intrinsic physical parameter and therefore their
study may lead to an enhanced understanding of AGN in general.

This proceedings contribution has two parts. First I describe and
discuss some of the  results obtained from a study of the X-ray
variability in NLS1s using {\it ASCA} data (Leighly 1999ab).  Then, I
discuss the results from our {\it HST} observations of NLS1s that were
first presented during this meeting  (Leighly \& Halpern 2000).
\end{abstract}

\begin{keyword}
 galaxies: active --- galaxies: individual (IRAS~13224$-$3809,
 1H~0707$-$495) --- line: profiles -- ultraviolet: galaxies -- X-rays:
 galaxies
\end{keyword}

\end{frontmatter}

% -------------------------------------------------------------------------

\newpage
\section{Introduction}

{\it ASCA} observations have proved to be instrumental in our
understanding of Narrow-line Seyfert 1 galaxies (NLS1s).  One of the
most compelling first indications that NLS1s may be characterized by a
high accretion rate came from the {\it ASCA} spectrum of RE~1034+39,
reported by Pounds, Done \& Osborne (1995).  This spectrum revealed a
strong soft excess component and a steep hard X-ray power law that
seemed to be similar to the spectra of black hole candidates in the
high state.  While it is now quite clear that not all NLS1s have such
spectra and indeed a range of strengths of the soft excess are seen
(e.g.\ Leighly 1999b), and that perhaps comparison with the {\it very}
high state seen in some Galactic X-ray emitting objects may be more
appropriate, the result has been extremely important in the
development of our understanding of these objects.

The results from the {\it ASCA} observations of NLS1s have now been
reported in several places (e.g.\ Leighly 1999ab and references
therein, see also Vaughn et al.\ these proceedings). Therefore, only
part of this review will be devoted to a few results from the
variability analysis of NLS1s and a very brief discussion of models and
implications.  In the second half, I present a preview of our work on
the {\it HST} spectra from NLS1s (Leighly \& Halpern 2000).

\section{X-ray Variability of NLS1s}

The X-ray variability in NLS1 {\it ASCA} observations is discussed in
detail in Leighly 1999a.  Because of the gaps in the light curves, the
{\it excess variance}, also known as the fractional amplitude of
variability, was used to quantify the variability.  The results are
shown in Fig.~1 (left).  This figure shows that for a given X-ray
luminosity, the amplitude of variability is consistently larger for
NLS1s than for Seyfert 1 galaxies with broad optical lines (BLS1s).
The uniform sampling and other properties of the {\it ASCA} data mean
that the excess variance should be proportional to the inverse of a
time scale of variability $T_i$ as $\propto T_i^{1-\alpha}$, where
$\alpha$ is the slope of the variability power spectrum (see also
Lawrence \& Papadakis 1993).  Thus, the simplest interpretation of
Fig.~1 is that, if the hard X-ray luminosity is characteristic of the
absolute mass accretion rate (i.e.\ the efficiency of conversion of
accretion energy to radiation is the same in all objects), then NLS1s
have a shorter variability time scale than do Seyfert 1s with broad
optical lines (BLS1s). This would be equivalent to compressing
a BLS1 light curve into an interval approximately 10 times shorter in
NLS1s. The variability time scale may be characteristic of the black
hole mass, simply from light-crossing time-scale arguments.  Thus, the
simplest explanation of this result is that NLS1s have a smaller black
hole mass than BLS1s, and since they are no fainter than BLS1s, they
must be accreting at a relatively larger accretion rate.

\begin{figure}[htb]
\centerline{\psfig{figure=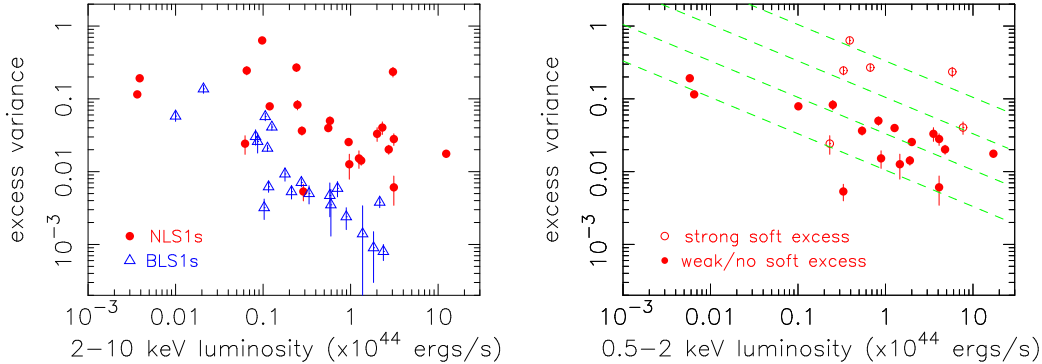,height=1.9truein,angle=270}}
\caption{{\it left:} Excess variance as a function of 2--10 keV luminosity
from a sample of NLS1s (Leighly 1999a) and broad-line Seyfert 1s
(Nandra et al.\ 1997).  For a particular hard X-ray luminosity, the
NLS1s show about an order of magnitude higher amplitude of
variability.  {\it right:}  Excess variance for NLS1s, here as a
function of 0.5--2.0 keV intrinsic luminosity.}
\end{figure}

Closer examination indicates that the situation may not be as simple
as the argument above would imply.  Specifically, Fig.~1 shows a
large spread in the excess variance values for NLS1s, much larger than
the measurement error, but also much larger than the model-dependent
systematic error in $\log$(excess variance) of $\sim$0.31--0.47
arising from the weakly nonstationary nature of a $1/f^\alpha$ power
spectrum (see Leighly 1999a for details).  It is possible that the
large spread is a consequence of the fact that the nature of the
variability may not be homogeneous among NLS1s.  Leighly 1999b report
a correlation between the strength of the soft excess in the {\it
ASCA} spectrum and the amplitude of the variability, and suggests that
strong soft excess objects are characterized by flaring, high
amplitude variability.  In Fig.~1 (right), I separate the 6 objects
identified as having strong soft excesses in Leighly 1999b and plot
the excess variance as a function of 0.5--2.0 keV
luminosity\footnote{Note that this is the inferred intrinsic
luminosity and therefore there may be some model dependence associated
with the modeling of the absorption.}.  When these six objects are
excluded, the slope of the regression at $-0.3$ is no longer biased by
the scatter.  The slope is still somewhat flatter than that expected
if the time scale is proportional to the luminosity assuming that the
slope of the variability power spectrum is $\alpha=1.5$ (dashed
lines)\footnote{Lawrence \& Papadakis (1993) found that the slopes of the
power spectra on 1-day time scales in a sample of objects are
consistent with a constant value of 1.55.}.

What is the origin of the enhanced variance observed in NLS1s?  As
stated above, it may be simply a consequence of a smaller emission
region characteristic of a smaller black hole.  However, the light
curves in some strong soft excess NLS1s are characterized by very high
amplitude flares, and they could not be simply time-compressed
versions of BLS1 light curves.  A few other emission mechanisms have
been suggested that instead would be equivalent to stretching the
amplitude of flares in BLS1 lightcurves to produce the enhanced excess
variance observed in NLS1s.  Naturally, because of the scale invariant
nature of the variability power spectrum (at least in the range of
frequencies that the {\it ASCA} data probe) these general scenarios
cannot be distinguished using the power spectrum or the excess variance.
\begin{itemize}
\item Noting that the very large amplitude variability on short time
scales implies a very high efficiency of conversion of accretion
energy to radiation in the NLS1 PHL~1092, it has been proposed that
beaming plays an important role in amplifying flares (Brandt et al.\
1999).  A specific scenario has been proposed: the emission regions
are found on the very inner edge of the accretion disk, and the
accretion disk is inclined at a very high angle with respect to the
viewer so that Doppler effects cause amplification of flares (Boller
et al.\ 1997).  The results presented in the next section of this
contribution cast significant doubt on an edge-on orientation.
\item It has recently been suggested that if NLS1s are accreting at a
higher rate then the magnetic field energy, assumed to be in
equipartion with gravitational potential energy, should be
proportionally larger (Mineshige et al.\ 2000).  Large amplitude
flares may result from reconnection of this more powerful magnetic
field.
\item It has been suggested that occultations by large optically-thick
clouds may produce high-amplitude X-ray variability (Brandt et al.\
these proceedings).  An observed steep $\alpha_{ox}$ in some objects
seems to support this idea; however, there is no physical reason that
luminous NLS1s should not have intrinsically steep $\alpha_{ox}$, for
example,  if the X-ray emitting corona is weak or not present.
\end{itemize}

What are the implications of the high amplitude flaring? It has been
previously suggested that high amplitude flaring is evidence that the
variability is nonlinear (Green, McHardy \& Done 1999; Boller et al.\
1997).  This conclusion is based on the assumption that the AGN
variability is Gaussian, since in that case a very high variance
compared with the mean can only be produced if the variability is
nonlinear.  In fact, non-Gaussian variability makes more sense for
AGNs because our usual physical picture supposes that the light curve
should be built up from the superposition of flares, and flares are
inherently non-Gaussian.  Linear, non-Gaussian variability can easily
produce high values of excess variance.  Non-Gaussian variability can
be detected using a parameter related to the skew of the flux
distribution, and evidence for non-Gaussianity was found in several of
the NLS1s with the highest amplitude of variability.  Demonstration
that a light curve is nonlinear is a much harder problem; it is a very
important one, however, because of the potentially strong constraints
on emission processes and geometry. See Leighly 1999a for a detailed
discussion.  We find some evidence for nonlinear variability in the
broad-line radio galaxy 3C~390.3 (Leighly \& O'Brien 1997) in that
quiescent periods occur before and after large flares, and this
behavior cannot be reproduced by a linear non-Gaussian model.
Evidence for nonlinear variability has been recently reported in
Cyg~X-1 (Timmer et al.\ 2000).

\section{{\it HST} Observations of NLS1s IRAS 13224$-$3809 and
1H~0707$-$495}  

In 1997, we reported the detection of an absorption feature near 1 keV
in the {\it ASCA} spectra from three NLS1s, IRAS~13224$-$3809,
1H~0707$-$495, and PG~1404+226 (Leighly et al.\ 1997).  Absorption by
ionized oxygen is common in the X-ray spectra from Seyfert 1 galaxies;
however, if we are to interpret the 1~keV features in this way, the
absorbing material must have highly relativistic velocities
(0.2--0.6$c$, depending on whether the features are interpreted as
absorption lines or edges).  Acceleration of ionized gas to these high
velocities should be difficult; however, we noted that NLS1s exhibit
suggestive similarities to a subclass of Broad-Absorption Line Quasars
(BALQSOs), objects characterized by the signature of absorption by
outflowing gas in the UV: both types of objects exhibit strong or
extreme \ion{Fe}{II} and weak [\ion{O}{III}] emission, they often have
red continua and strong infrared emission, and they are predominately
radio-quiet (see Leighly et al.\ 1997 for details).  An alternative
explanation as absorption by highly ionized neon and iron has been
presented by Nicastro et al.\ (1999).  With luck, the origin of these
features will be resolved by our upcoming {\it Chandra} observation of
1H~0707$-$495.

\begin{figure}[htb]
\centerline{\psfig{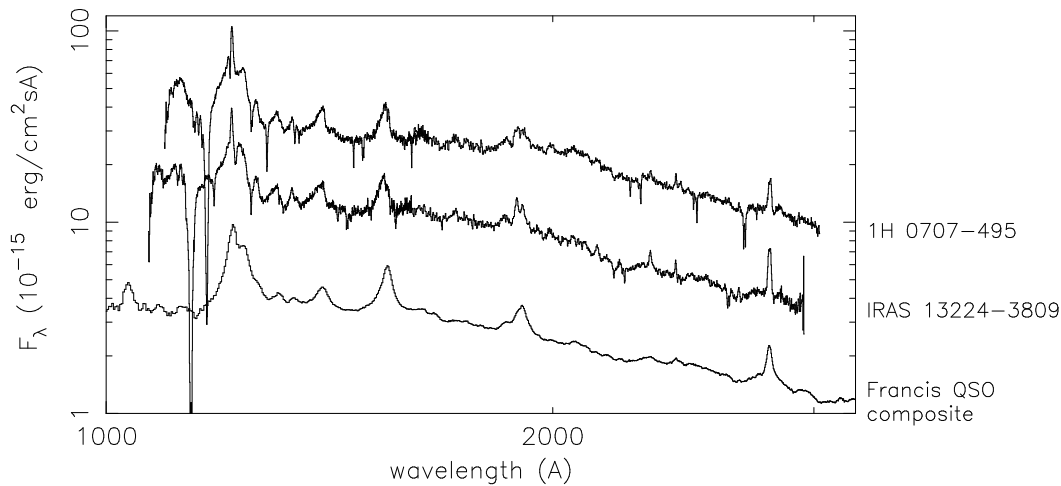}}
\caption{The broad-band UV continuum spectra from {\it HST}
observations of two NLS1s (IRAS~13224$-$3809 and 1H~0707$-$495)
plotted on a log--log scale and compared to the average quasar
spectrum obtained by Paul Francis.  The {\it HST} data have been
corrected for Galactic reddening. This plot demonstrates that the
NLS1 continua are as blue as that of the average quasar.}
\end{figure}

Absorption in X-rays is often accompanied by absorption in the UV.  To
test our hypothesis that the X-ray absorption features are produced by
relativistically outflowing gas, we applied for and were awarded {\it
HST} STIS UV spectroscopic observations of IRAS~13224$-$3809 and
1H~0707$-$495 (Leighly \& Halpern 2000) and the results of the
observations were presented for the first time at this meeting.  The
broad-band continuum spectra, corrected for Galactic reddening, are
presented in Fig.\ 2  along with the average QSO spectrum compiled by
Paul Francis. This figure shows that these NLS1s have continua 
as blue as the average QSO\footnote{IRAS~13224$-$3809 has been
previously reported as having a red spectrum (Mas-Hesse et al.\ 1994).
We find that our optical spectrum obtained from a 1 hour exposure on
the CTIO 4 meter telescope is indeed rather red and does not join
smoothly to the UV spectrum.  We infer, from a difference in the
spatial profile of the lines and continuum, that
there is a strong, red galaxy component in the spectrum.  In contrast,
our 1H~0707$-$495 optical spectrum is quite blue and joins smoothly onto the
UV spectrum.}.  There is also no evidence for resonance-line absorption
intrinsic to the object; all of the narrow absorption lines in the 
spectra originate in our Galaxy.

\begin{figure}[htb]
\centerline{\psfig{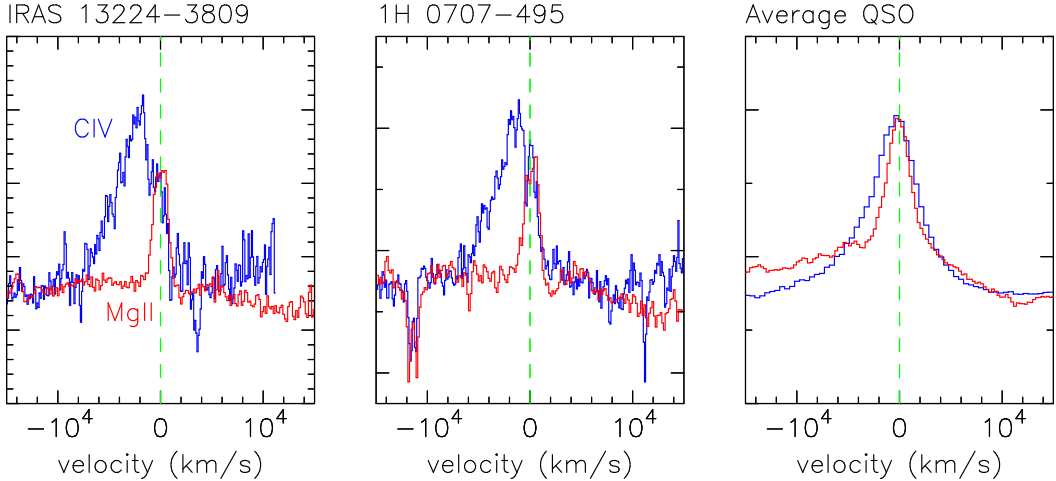}}
\caption{The rescaled representative high-ionization line \ion{C}{IV}
superimposed on the representative low-ionization line \ion{Mg}{II} as
a function of velocity, for our two NLS1s and the average quasar.  The
average quasar high-ionization line is slightly broader and slightly
blueshifted compared with the low-ionization line.  In contrast, the
low-ionization lines in NLS1s are much narrower and the
high-ionization lines are strongly blueshifted.}
\end{figure}

The emission line profiles are the most interesting feature of these
spectra.  The high-ionization lines, including Ly$\alpha$$\lambda
1216$, \ion{N}{V}$\lambda 1240$, \ion{Si}{IV}$\lambda 1397$ and
\ion{C}{IV}$\lambda 1549$ have a different profile than the
low-ionization lines, including \ion{Mg}{II}$\lambda 2800$ and
H$\beta$$\lambda 4861$.  The difference is illustrated in Fig.\ 3
which overlays the rescaled high-ionization line \ion{C}{IV} and the
low-ionization line \ion{Mg}{II} profiles as a function of
velocity\footnote{Note that the profile of the doublet
\ion{Mg}{II}$\lambda$2796, $\lambda$2804 is consistent with that of
H$\beta$.}.  As is commonly found in AGN (e.g.\ Marziani et al.\ 1997),
the average quasar high-ionization lines are broader and blueshifted
relative to the low-ionization lines.  The same trend is true in
NLS1s, except the difference is more profound. The \ion{C}{IV} FWHM is
$4 \times$ that of H$\beta$, but it is significant that all of the
extra width is on the blue side, whereas the red side lines up well
with the low-ionization lines.  Similar line profiles were reported
from the NLS1 I~Zw~1 (Laor et al.\ 1997).

\begin{figure}[htb]
\centerline{\psfig{figure=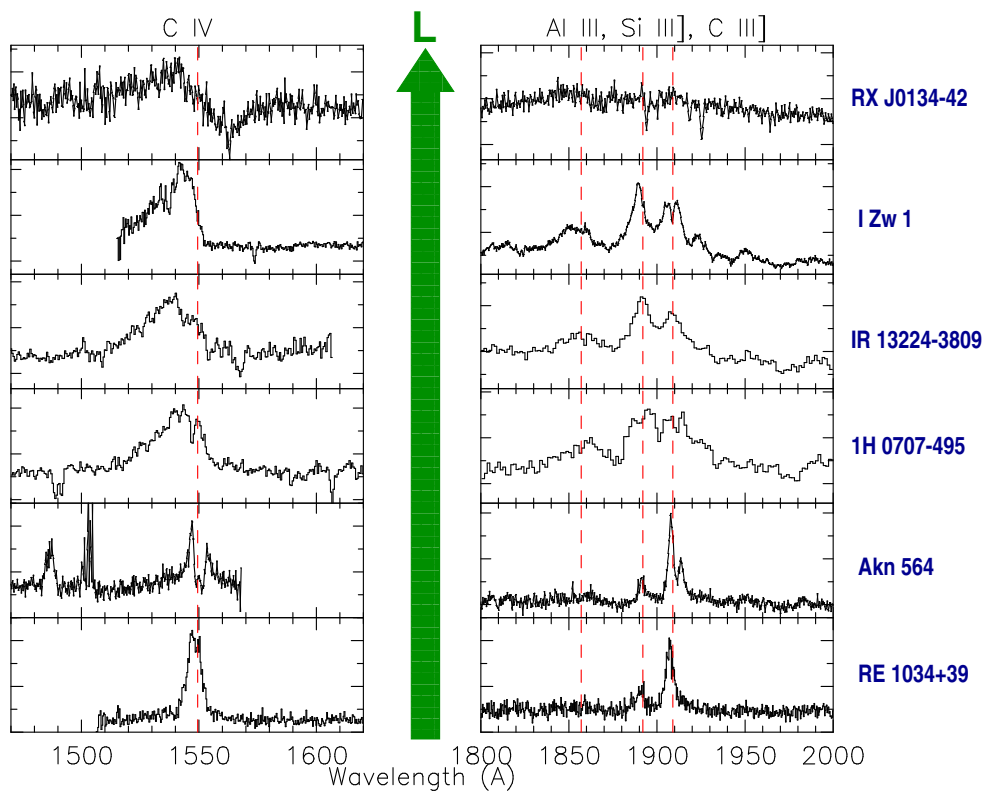,height=4.1truein,angle=270}}
\caption{Spectra near CIV$\lambda 1549$, and near AlIII$\lambda 1857$,
SiIII]$\lambda 1892$ and CIII]$\lambda 1909$ as a function of
arbitrarily rescaled flux, and in order of decreasing luminosity.
Blueshifted, broad CIV lines and high SiIII] to CIII] ratios are
present in the higher-luminosity objects.}
\end{figure}

We also find evidence for emission from gas with very high densities.
In Fig.\ 4 we plot the \ion{Al}{III}--\ion{Si}{III}]--\ion{C}{III}]
line region of our spectra, as well as archived spectra from several
other NLS1s.  We find a very high \ion{Si}{III}] to \ion{C}{III}]
ratio in our spectra from IRAS~13224$-$3809 and 1H~0707$-$495.  The
ions responsible for these lines are found under the same physical
conditions; however, \ion{Si}{III}] has a higher critical density than
\ion{C}{III}] and thus the ratio becomes large when the \ion{C}{III}]
emission has saturated.  A correlation between this ratio and the
width of H$\beta$ has been previously reported by Wills et al.\ 1999
(see also Wills, these proceedings). We note that these emission lines
are relatively narrow and symmetric about their rest wavelengths and
thus their profiles are more similar to those of the low-ionization
lines.

Fig.\ 4 compares line profiles from 6 NLS1s. Three of the objects
(I~Zw~1, IRAS~13224$-$3809 and 1H~0707$-$495) display blueshifted
high-ionization lines and high \ion{Si}{III}] to \ion{C}{III}] ratios.
RX~J0134$-$42 also appears to have a blueshifted \ion{C}{IV} line.
However, the spectra from two lower luminosity NLS1s, Akn~564 and
RE~1034+39, lack both of these attributes.  We suspect that this
reflects a dependence of the line properties on luminosity.

\subsection{Disk-wind Models of AGN}

We propose that the characteristic spectra we observe in the higher
luminosity NLS1s can be naturally explained by a disk-wind model, as
shown schematically in Fig.\ 5.  It has long been known that it is not
possible to produce all broad-line emission in gas characterized by
one set of physical properties.  Several multicomponent models,
including disk-wind models, have been proposed; unfortunately, space
limitations prohibit a through review here.  What is new here is that
our observations provide evidence that a disk-wind model is strongly
favored over all other models.  The fact that the red side of the high-
and low-ionization lines line up suggests that the low-ionization
lines and low-velocity cores of the high-ionization lines are produced
in the same relatively low-velocity material (i.e.\ the accretion
disk) while the high-velocity blue wings of the lines are produced in
a wind.  The disk is optically thick, and therefore, we see only
emission from the wind accelerated toward us.

\begin{figure}[htb]
\centerline{
\psfig{bbllx=227pt,bblly=191pt,bburx=385pt,bbury=600pt,figure=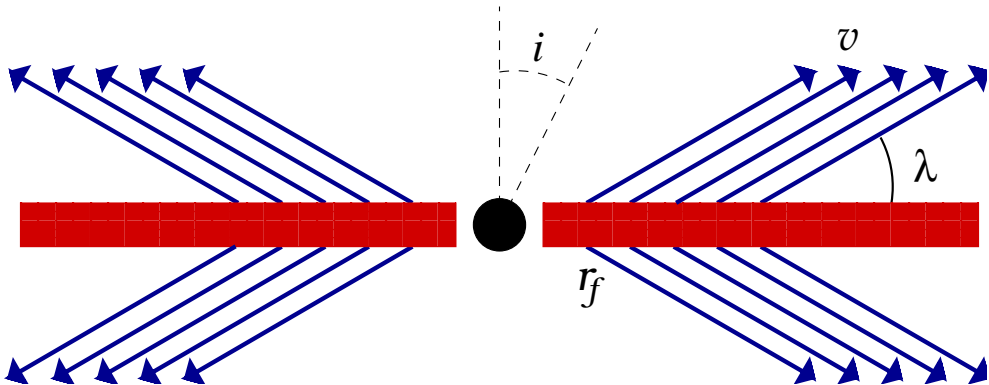,height=2.0truein,clip=1,angle=270}}
\caption{Schematic diagram of our simple disk-wind model.  The
system is observed at an inclination angle $i$, the wind stream lines
make an angle $\lambda$ with the disk, and the flow is initiated at
the foot-point radius $r_f$.}
\end{figure}

We are currently constructing a simple disk-wind model to test our
assertion and Fig.\ 5 shows a schematic of the assumed geometry.  We
are following Murray \& Chiang (1998) in general: we use a $\beta$
velocity law, often applied in star and CV winds, and we are using
{\it Cloudy} to estimate the ionization structure.  However, since we
are interested in the emission produced in the wind in particular, we
treat the radiative transfer numerically using the Sobolev
approximation.  A complementary and more sophisticated model being
constructed by Proga, Stone and Kallman (2000; hereafter PSK) aims to
determine the detailed dynamics of the wind.  A definitive model has
not yet been constructed; however, it is clear that there are
potentially complex relationships between the radiative transfer,
dynamics and ionization, and the results will quite possibly not
conform to naive intuition when everything has been taken into
account.  Examples of two complications:
\begin{itemize}
\item The wind is quite likely to be optically thick to scattering of
the resonance lines, and thus radiative transfer will be strongly
enhanced along directions in which the velocity gradient is high.
This effect can strongly alter the observed line profiles for the
high-ionization lines emerging from the wind, and prompts the use of
the Sobolev approximation.  The \ion{He}{II} recombination lines are
immune to this effect, however.
\item The acceleration of the wind is produced by resonance-line
driving and thus the velocities attained depend intimately on the
ionization structure of the gas.  Specifically, there must be a
mechanism for shielding the wind gas from the photoionizing X-ray
source.  This problem seems to have a natural solution: PSK find that
there can be adequate ``failed'' wind material falling back onto the
object to do this.
\end{itemize}

Our preliminary results and also the results presented by PSK are
promising enough to conjecture that many of the features of the {\it HST}
NLS1 spectra will be explainable using a disk-wind system. For example:
\begin{itemize}
\item A disk-wind model coupled with a high accretion rate may
naturally explain NLS1 spectra.  PSK find that a
primary condition on formation of the wind is that the wind-driving
(UV) flux should be large compared with the wind-photoionizing
(X-ray) flux.  A high accretion rate predicts a 
spectrum strongly peaked in the extreme UV and therefore the UV to
X-ray ratio will be large enough that a strong wind can be driven.  If
the accretion rate is high, then densities in the disk will be high,
resulting in a large \ion{Si}{III}] to \ion{C}{III}] ratio.  More
material will be available to be blown from the nucleus, and since the
black hole mass is relatively smaller, the wind is more likely to
reach escape velocity.  The wind may partially shield the accretion
disk from the intense radiation field, thus permitting excitation but
not causing ionization of the Fe$^+$ ion,  so that strong optical
\ion{Fe}{II} emission may be produced.  This could explain the
observed correlation between high-ionization blue asymmetry and
optical \ion{Fe}{II} (e.g.\ Marziani et al.\ 1996).  
\item The large UV to X-ray ratio requirement found by PSK may explain
why lower luminosity NLS1s do not show blue wings on their
high-ionization lines.  Quasars are known to have typically relatively
larger UV to X-ray ratios than Seyferts (Wilkes et al.\ 1994), and thus
gas in Seyferts may become too highly ionized to be accelerated to
high velocities.
\item The observed line profiles will be affected by the observer's
viewing angle, and radiative transfer effects, as noted above,
potentially affect the profiles as well in a non-obvious way.  We
interpret the fact that the red sides of the high- and low-ionizations
lines line up as evidence that the low-ionization lines and the
low-velocity core of the high-ionization lines are produced in the
disk. However, for there to be only blue-side wind emission, either
the observer's viewing angle must be nearly face on, or the angle of
the wind stream lines with respect to the disk must be large.  It is
quite possible that in reality the wind stream line angle and the
radius at which the wind is launched are not independent of the
accretion rate, and thus results from a self-consistent model would be
potentially very interesting.  
\end{itemize}

\subsection{Summary and Further Implications}

We have described how the particular UV spectra from higher luminosity
NLS1s strongly suggests that the emission lines are produced in a
disk-wind system. Further, we conjecture that, through a potentially
complicated dependence of model parameters such as the amount of
material and velocity and angle of stream lines on the physical
parameters such as the black hole mass and accretion rate, it may be
possible to explain Eigenvector 1 self-consistently.
There are a few more important issues:
\begin{itemize}
\item Broad emission lines in AGN are often assumed to be produced in
clouds that have virialized bulk motions (e.g.\ Peterson \& Wandel
1999).  In contrast, the strong winds that we propose are responsible
for the blue wings in our spectra and thus for a significant fraction of
the high-ionization line emission in luminous NLS1s will not have
virialized bulk motion.
\item It has been proposed that NLS1s bear some similarity to high
redshift quasars (Mathur 2000): in particular, very high redshift
quasars ($z \approx 4$) appear to have rather narrow lines compared
with local quasars (Shields et al.\ 1997).  The spectra presented here
show that this conjecture may have serious weaknesses.  At high
redshift, only the rest-frame UV high-ionization lines are observed in
the optical spectrum.  We have discovered that the high-ionization
lines in higher-luminosity NLS1s are typically broad; furthermore,
Wills et al.\ (these proceedings) find no correlation between the
high-ionization and H$\beta$ line widths in the PG-quasar subsample.
These results imply that the observed optical spectra of 
high-redshift NLS1s will show {\it broad} emission lines, and
therefore, such objects may be difficult to identify. Finally, if
there is a wind, it is possible that some of the \ion{N}{V} emission
is a result of excitation of high velocity N$^{+4}$ ions by absorption
of Ly$\alpha$ photons, and abundance enhancements may be less
necessary (however, see also Hamann \& Korista 1996).
\item Will the wind be homogeneous? Resonance-line driven winds are a
characteristic of Wolf-Rayet stars and there is very strong evidence
for density enhancements in these winds. Furthermore, PSK find density
inhomogeneities in their simulations.  Clumping may be necessary to
match the intensity of observed emission lines.  If present, clumping
could produce interesting time-dependent and ionization-dependent
behavior.
\end{itemize}

\par\noindent
{\bf Acknowledgements:} I would like to thank the Wilhelm and Else
Heraeus Foundation for travel support, Jules Halpern for helpful
discussions, and Daniel Proga for an early look at his submitted paper.
I gratefully acknowledge support through NAG5-7971 (NASA LTSA).

% -------------------------------------------------------------------------

% -------------------------------------------------------------------------

\end{document}